\pdfoutput=1
\documentclass[sigconf,screen,nonacm]{acmart}
\AtBeginDocument{}

\usepackage{listings}
\usepackage{booktabs}
\usepackage{subcaption}
\usepackage{xspace}
\usepackage{tikz}
\usepackage{colortbl}
\usetikzlibrary{positioning,arrows.meta,calc}

\newcommand{\gensparql}{\textsc{GenSPARQL}\xspace}
\newcommand{\genop}{\texttt{GENOP}\xspace}

\newcommand{\code}[1]{\texttt{#1}}
\newcommand{\sem}[1]{\mathopen{[\![}#1\mathclose{]\!]}}

\definecolor{gsKeyword}{RGB}{0,90,160}
\definecolor{kgfill}{HTML}{DDEBF8}
\definecolor{kgdraw}{HTML}{77AADD}
\definecolor{kgtext}{HTML}{4477BB}
\definecolor{genfill}{HTML}{F8F1CF}
\definecolor{gendraw}{HTML}{DDCC77}
\definecolor{tolblue}{HTML}{4477BB}
\definecolor{tolgreen}{HTML}{228833}
\definecolor{tolpurple}{HTML}{882288}
\definecolor{tolpaleblue}{HTML}{AACCEE}
\definecolor{tolpalegreen}{HTML}{BBDDBB}
\definecolor{tolpalepurple}{HTML}{EEBBDD}
\definecolor{gsComment}{RGB}{110,110,110}
\definecolor{gsString}{RGB}{160,60,20}

\lstdefinelanguage{gensparql}{
  morekeywords={SELECT,WHERE,PREFIX,FILTER,OPTIONAL,UNION,BIND,GENOP,DISTINCT,ORDER,BY,LIMIT,GROUP},
  sensitive=true,
  morecomment=[l]{\#},
  morestring=[b]",
}
\begin{document}

\title{GenSPARQL: A Generative Extension of SPARQL for Incomplete Knowledge Graphs}

\author{Jingcheng Wu}
\email{jingcheng.wu@ki.uni-stuttgart.de}
\affiliation{\institution{University of Stuttgart}\city{Stuttgart}\country{Germany}}

\author{Yi Wang}
\email{yi.wang@ki.uni-stuttgart.de}
\affiliation{\institution{University of Stuttgart}\city{Stuttgart}\country{Germany}}

\author{Hongkuan Zhou}
\additionalaffiliation{\institution{Bosch Corporate Research}\city{Renningen}\country{Germany}}
\email{hongkuan.zhou@de.bosch.com}
\affiliation{\institution{University of Stuttgart}\city{Stuttgart}\country{Germany}}

\author{Steffen Staab}
\additionalaffiliation{\institution{University of Southampton}\city{Southampton}\country{UK}}
\email{steffen.staab@ki.uni-stuttgart.de}
\affiliation{\institution{University of Stuttgart}\city{Stuttgart}\country{Germany}}

\author{Bo Xiong}
\email{boxiong@iastate.edu}
\affiliation{\institution{Iowa State University}\city{Ames}\state{IA}\country{USA}}

\begin{abstract}
Knowledge graphs on the Web are incomplete, and SPARQL answers a query only from the stored triples. Large language models (LLMs) can predict missing facts, but an LLM returns names instead of IRIs and requires one call per binding of a query variable. We present \gensparql{}, a generative extension of SPARQL~1.1 with an operator \genop{} that binds values generated by an LLM to query variables during query evaluation. A similarity join maps generated names to entities of the graph, and the query planner invokes the LLM only for the bindings selected by the other graph patterns and filters. In a web workbench, attendees query two graphs built from Wikidata and inspect generated values, mapped entities and LLM calls. In our evaluation with three open-weight LLMs, the embedding similarity maps 83\% of the entities named by the LLMs and yields only one incorrect match.
\end{abstract}




\maketitle

\section{Introduction}
\label{sec:intro}

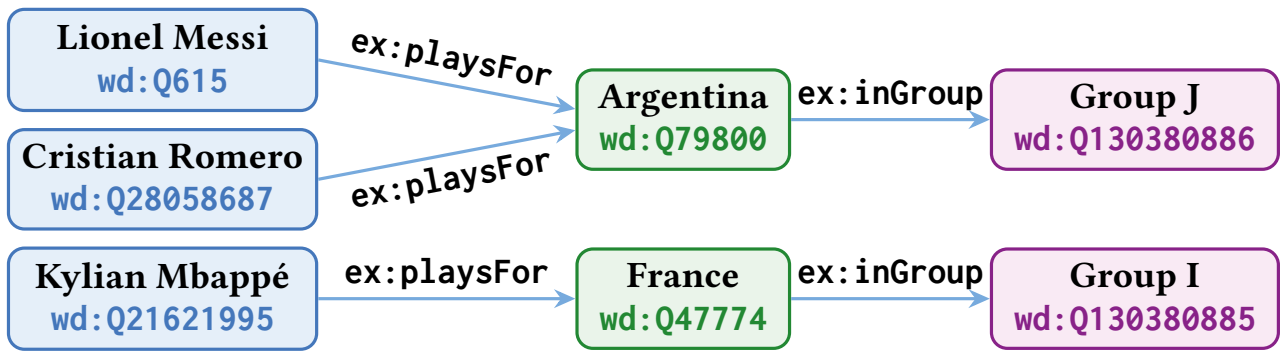
\begin{figure}[t]
\centering
\begin{tikzpicture}[x=1pt, y=1pt, font=\footnotesize\bfseries,
  ent/.style={rounded corners=3pt, align=center, inner sep=2.5pt, line width=0.6pt, fill opacity=0.3, text opacity=1},
  pl/.style={ent, draw=tolblue, fill=tolpaleblue},
  tm/.style={ent, draw=tolgreen, fill=tolpalegreen},
  gr/.style={ent, draw=tolpurple, fill=tolpalepurple},
  rel/.style={-{Stealth[length=4pt,width=3.5pt]}, draw=kgdraw, line width=0.6pt},
  rl/.style={font=\ttfamily\footnotesize\bfseries, inner sep=1.5pt}]
\newcommand{\iri}[2]{{\ttfamily\color{#1}#2}}
\node[pl, minimum width=58pt] (messi) at (0,0) {Lionel Messi\\\iri{tolblue}{wd:Q615}};
\node[pl, minimum width=58pt, below=3pt of messi] (romero) {Cristian Romero\\\iri{tolblue}{wd:Q28058687}};
\node[pl, minimum width=58pt, below=3pt of romero] (mbappe) {Kylian Mbapp\'e\\\iri{tolblue}{wd:Q21621995}};
\node[tm, minimum width=40pt] (arg) at ($(messi)!0.5!(romero)+(98,0)$) {Argentina\\\iri{tolgreen}{wd:Q79800}};
\node[tm, minimum width=40pt] (fra) at ($(mbappe)+(98,0)$) {France\\\iri{tolgreen}{wd:Q47774}};
\node[gr, minimum width=54pt] (gj) at ($(arg)+(85,0)$) {Group J\\\iri{tolpurple}{wd:Q130380886}};
\node[gr, minimum width=54pt] (gi) at ($(fra)+(85,0)$) {Group I\\\iri{tolpurple}{wd:Q130380885}};
\draw[rel] (messi.east) -- node[rl, sloped, above] {ex:playsFor} ([yshift=2pt]arg.west);
\draw[rel] (romero.east) -- node[rl, sloped, below] {ex:playsFor} ([yshift=-2pt]arg.west);
\draw[rel] (mbappe.east) -- node[rl, above] {ex:playsFor} (fra.west);
\draw[rel] (arg.east) -- node[rl, above] {ex:inGroup} (gj.west);
\draw[rel] (fra.east) -- node[rl, above] {ex:inGroup} (gi.west);
\end{tikzpicture}
\caption{Excerpt of the 2026 World Cup graph.}
\label{fig:kg}
\end{figure}

Knowledge graphs on the Web, such as Wikidata~\cite{vrandecic2014wikidata}, are incomplete~\cite{hogan2021kg}. SPARQL is the W3C-recommended query language for knowledge graphs in RDF~\cite{sparql11query}. A SPARQL query matches its graph patterns only against the stored triples~\cite{sparql11}. For example, Figure~\ref{fig:kg} shows an excerpt of a knowledge graph of the 2026 FIFA World Cup with teams, groups and players from Wikidata. The query
\par\vspace{3pt}\noindent\resizebox{\columnwidth}{!}{\texttt{\textcolor{gsKeyword}{\textbf{SELECT}} ?pos \textcolor{gsKeyword}{\textbf{WHERE}} \{ ?p ex:playsFor wd:Q79800 ; ex:position ?pos \}}}\par\vspace{3pt}\noindent
asks for the positions of the Argentine players and returns an empty result, because the graph models no playing positions.

Large language models (LLMs) acquire factual knowledge during pretraining on Web documents such as Wikipedia articles~\cite{petroni2019lama} and generate candidate answers to questions about facts that a graph lacks, such as the positions and the World Cup titles of players. Extending the answers of a SPARQL query with values generated by an LLM poses three obstacles. First, an LLM returns entity names as strings, whereas the graph identifies entities by IRIs, such as \code{wd:Q615} for Lionel Messi. A SPARQL graph pattern relates a generated name to an entity only through an identical label and therefore cannot match variants such as Emiliano Martinez to the label Emiliano Mart\'inez. Second, an LLM also generates names of entities outside the graph and names that denote no entity at all~\cite{kghalluc}. A name outside the graph can resemble the label of a different entity, so a comparison that tolerates variants can also produce incorrect matches. Third, a prompt with a query variable yields one LLM call per binding, such as one call per player.

Four lines of existing work combine LLMs with queries over knowledge graphs. Text-to-SPARQL systems employ an LLM to translate a question in natural language into a SPARQL query~\cite{llmsparql}, whose answers remain restricted to the stored facts. Work on LLM-augmented knowledge graphs~\cite{llmkg} materializes facts inferred by an LLM in the graph before query evaluation and therefore spends LLM calls on all entities of the graph, including entities that no query reads. Retrieval-augmented generation over knowledge graphs~\cite{graphrag} retrieves facts from the graph as context for an LLM and returns the answer of the LLM as text with names instead of IRIs, so a SPARQL query cannot use the named entities in further graph patterns. GraphDB invokes an LLM inside a SPARQL query through magic predicates and binds the generated text to a variable as a literal~\cite{graphdb}. The query itself has to relate the literal to an entity of the graph. The mapping of generated names to entities of the graph thus remains outside the query engine.

We present \gensparql{}, a generative extension of SPARQL. \gensparql{} extends the SPARQL algebra with the operator \genop{}, which invokes an LLM during query evaluation and binds the generated values to query variables. Within the query engine, a similarity join~\cite{simjoin} can map generated names to entities of the graph. A name matches an entity if the similarity between the name and a label of the entity reaches a threshold~$\theta$. The answer then contains the IRIs of the matched entities instead of the names, and the IRIs are joined with the solutions of the other graph patterns. Moreover, the query planner schedules \genop{} after the triple patterns that contain no variable bound by \genop{} and after the filters over variables bound by the same triple patterns. The LLM is therefore invoked only for the solutions of the triple patterns and filters.

The demonstration applies \gensparql{} to the World Cup graph and a graph of Nobel laureates from Wikidata. In a web workbench (Section~\ref{sec:demo}), attendees write \genop{} queries, choose the LLM and the threshold~$\theta$, and inspect the matched entities and the number of LLM calls of each query.\footnote{Code (MIT license): \url{https://github.com/0sidewalkenforcer0/GenSPARQL}.\\ Project page: \url{https://0sidewalkenforcer0.github.io/GenSPARQL/}.\\ Video: \url{https://0sidewalkenforcer0.github.io/GenSPARQL/video/gensparql-demo.mp4}.} A recorded mode on the project page requires no LLM.
\section{GenSPARQL}
\label{sec:gensparql}

In this section, we define the operator \genop{} (Section~\ref{sec:genop}) and the similarity join that maps generated names to entities of the graph (Section~\ref{sec:grounding}) and describe the query planner and the implementation in Apache Jena (Section~\ref{sec:implementation}).

\begin{figure}[t]
\centering
\begin{tikzpicture}[x=1pt, y=1pt, font=\footnotesize\bfseries,
  ent/.style={draw=tolblue, fill=tolpaleblue, fill opacity=0.3, text opacity=1, rounded corners=2pt, inner sep=2pt, line width=0.6pt, font=\ttfamily\footnotesize\bfseries},
  gen/.style={draw=gendraw, fill=genfill, rounded corners=2pt, inner sep=2pt, line width=0.5pt},
  nm/.style={draw=tolblue, fill=tolpaleblue, fill opacity=0.3, text opacity=1, rounded corners=2pt, inner sep=2pt, line width=0.6pt},
  code/.style={draw=kgdraw!45, fill=white, rounded corners=3pt, align=left, inner sep=2.5pt, font=\ttfamily\footnotesize\bfseries, line width=0.5pt},
  map/.style={-{Stealth[length=3.5pt,width=3pt]}, draw=kgdraw, line width=0.5pt},
  sub/.style={font=\small\bfseries, anchor=west, inner sep=0pt}]
\newcommand{\kw}[1]{\textcolor{gsKeyword}{\textbf{#1}}}
\setlength{\fboxsep}{0.8pt}
\node[sub] (ta) at (0,0) {(a) Positions of the Argentine players};
\node[code, anchor=north west] (qa) at ([yshift=-3pt]ta.south west) {\kw{SELECT} ?name ?pos \kw{WHERE} \{\\
\ ?player ex:playsFor ?t .\\
\ ?t rdfs:label "Argentina" .\\
\ ?player rdfs:label ?name .\\
\ \colorbox{genfill}{\kw{GENOP}("Position of \{?name\}?",}\\
\ \colorbox{genfill}{\ \ (?pos), <model>)}\\
\}};
\node[anchor=north west, inner sep=0pt, font=\ttfamily\footnotesize\bfseries] (hn) at ([xshift=8pt,yshift=-1pt]qa.north east) {?name};
\node[nm, anchor=north west] (n1) at ([yshift=-3pt]hn.south west) {Lionel Messi};
\node[nm, anchor=north west] (n2) at ([yshift=-3pt]n1.south west) {Cristian Romero};
\node[nm, anchor=north west] (n3) at ([yshift=-3pt]n2.south west) {Emiliano Mart\'inez};
\node[gen, anchor=west] (v1) at ([xshift=4pt]n1.east-|n3.east) {Forward};
\node[gen, anchor=west] (v2) at ([xshift=4pt]n2.east-|n3.east) {Defender};
\node[gen, anchor=west] (v3) at ([xshift=4pt]n3.east-|n3.east) {Goalkeeper};
\node[anchor=base west, inner sep=0pt, font=\ttfamily\footnotesize\bfseries] at (hn.base west-|v1.west) {?pos};
\draw[line width=0.4pt] ([yshift=-1.5pt]hn.south west) -- ([yshift=-1.5pt]hn.south west -| v3.east);
\node[anchor=north, inner sep=0pt] (dn) at ([yshift=-1pt]n3.south) {\scalebox{0.7}{$\vdots$}};
\node[anchor=north, inner sep=0pt] (dv) at ([yshift=-1pt]v3.south) {\scalebox{0.7}{$\vdots$}};
\path let \p1=(qa.south), \p2=(dn.south) in coordinate (low) at (0,{min(\y1,\y2)});
\node[sub, anchor=north west] (tb) at ([yshift=-4pt]qa.south west |- low) {(b) Players who have won a World Cup};
\node[code, anchor=north west] (qb) at ([yshift=-3pt]tb.south west) {\kw{SELECT} ?player \kw{WHERE} \{\\
\ ?player a ex:Athlete .\\
\ \colorbox{genfill}{\kw{GENOP}("List footballers who}\\
\ \colorbox{genfill}{\ \ have won the World Cup.",}\\
\ \colorbox{genfill}{\ \ (?player), <model>, 0.85)}\\
\}};
\node[anchor=north west, inner sep=0pt, font=\ttfamily\footnotesize\bfseries] (hp) at ([xshift=8pt,yshift=-1pt]qb.north -| qa.east) {?player};
\node[ent, anchor=north west] (i1) at ([yshift=-3pt]hp.south west) {wd:Q615};
\node[ent, anchor=north west] (i2) at ([yshift=-3pt]i1.south west) {wd:Q21621995};
\draw[line width=0.4pt] ([yshift=-1.5pt]hp.south west) -- ([yshift=-1.5pt]hp.south west -| i2.east);
\node[gen, anchor=west] (g1) at ([xshift=12pt]i1.east-|i2.east) {Lionel Messi};
\node[gen, anchor=west] (g2) at ([xshift=12pt]i2.east) {Kylian Mbapp\'e};
\node[gen, anchor=north west] (g3) at ([yshift=-3pt]g2.south west) {Pel\'e};
\draw[map] (g1.west) -- (i1.east);
\draw[map] (g2.west) -- (i2.east);
\coordinate (legc) at ($(current bounding box.west |- ta.north)!0.5!(current bounding box.east |- ta.north)$);
\node[anchor=south, inner sep=0pt] at ([yshift=4pt]legc) {\tikz[baseline]{\node[draw=tolblue, fill=tolpaleblue, fill opacity=0.3, text opacity=1, rounded corners=2pt, inner sep=1.5pt, line width=0.5pt, font=\footnotesize\bfseries] (lb) {term of the graph}; \node[draw=gendraw, fill=genfill, rounded corners=2pt, inner sep=1.5pt, line width=0.5pt, font=\footnotesize\bfseries, anchor=west] at ([xshift=14pt]lb.east) {\genop{} or generated value};}};
\end{tikzpicture}
\caption{Two \genop{} queries over the 2026 World Cup graph.}
\label{fig:running}
\end{figure}
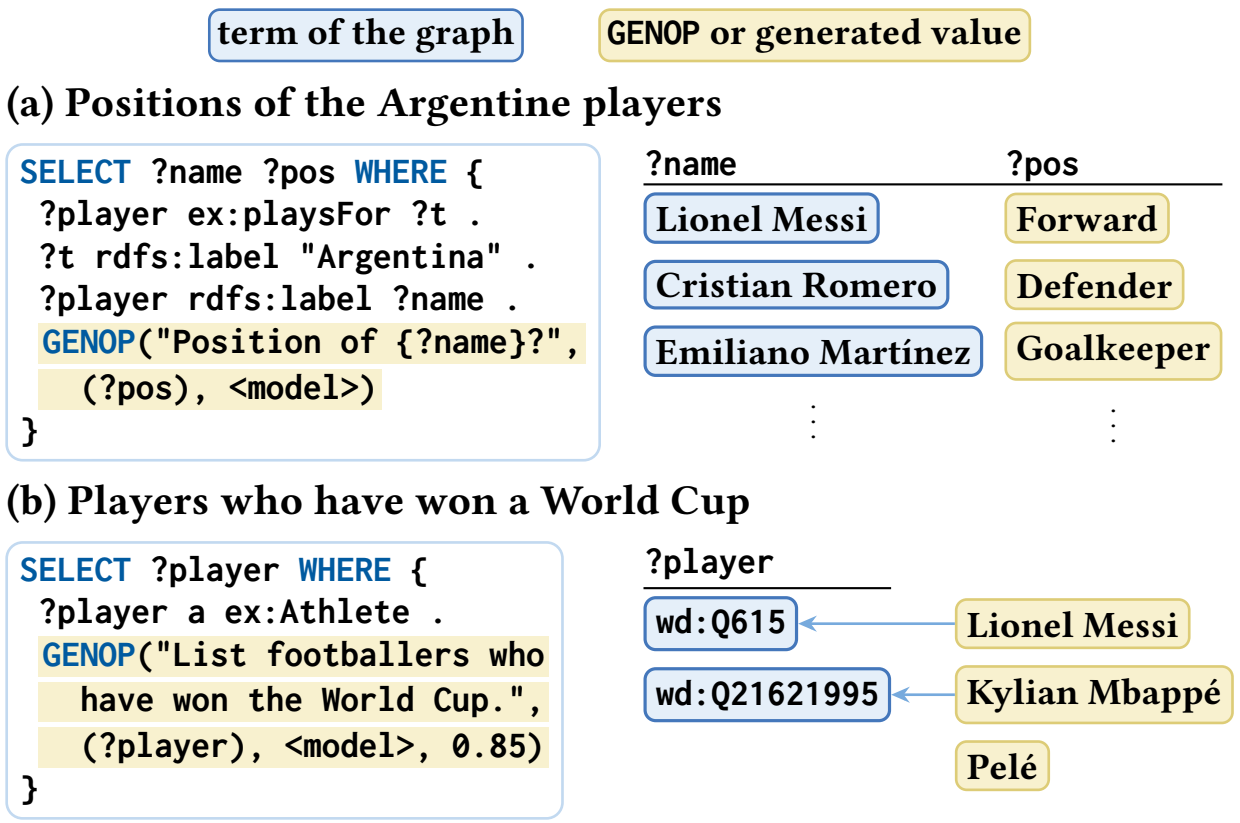

\subsection{\texorpdfstring{The \genop{} Operator}{The GENOP Operator}}
\label{sec:genop}

\paragraph{Syntax.}
We extend the grammar of SPARQL~1.1~\cite{sparql11query} with \genop{}. For a prompt string~$p$, output variables \code{?v1},~\dots,~\code{?vk}, an IRI~$m$ that identifies an LLM and a threshold~$\theta\in[0,1]$, the expression
\[
  \texttt{\textcolor{gsKeyword}{\textbf{GENOP}}(}p\texttt{, (?v1, ..., ?vk), }m\texttt{, }\theta\texttt{)}
\]
is an element of group graph patterns. The argument~$\theta$ is optional and defaults to a threshold~$\theta_0$. \genop{} expressions occur only outside \code{EXISTS} and \mbox{\code{NOT EXISTS}}. A placeholder \code{\{?x\}} in~$p$ refers to the query variable \code{?x}. A \genop{} expression is in \emph{context mode} if $p$ contains placeholders, and in \emph{base mode} otherwise.

\paragraph{Semantics.}
The algebra of \gensparql{} extends the SPARQL algebra~\cite{sparql11query} with \genop{} patterns and similarity joins. A \genop{} pattern has the form $\genop_\theta(p,\bar v,m,P)$, where $p$, $\bar v=(\mathord{?}v_1,\dots,\mathord{?}v_k)$, $m$ and~$\theta$ are the arguments of a \genop{} expression and $P$ is a graph pattern. \genop{} extends the solutions of~$P$ with values that the LLM~$m$ generates for~$\bar v$. In a query, $P$ is implicit and consists of the elements of the enclosing group that contain no variable of~$\bar v$. The elements that contain a variable of~$\bar v$ are joined with the extended solutions through a similarity join~\cite{simjoin} (Section~\ref{sec:grounding}).

In general, let $G_i=\genop(p_i,\bar v_i,m_i,\theta_i)$ for $1\le i\le n$ be the \genop{} expressions of a group graph pattern in written order. The placeholders of~$p_i$ can refer to the variables in $\bar v_1,\dots,\bar v_{i-1}$. Each triple pattern and each property path is a separate element of the group. For the set $\mathrm{var}(e)$ of variables of an element~$e$, let
\[
  \begin{aligned}
  Q_0 &= \{ e{}\notin R \,{:}\, \mathrm{var}(e)\cap(\bar v_1\cup\dots\cup\bar v_n)=\emptyset \},\\
  Q_i &= \{ e{}\notin R \,{:}\, \mathrm{var}(e)\cap\bar v_i\neq\emptyset,\ \mathrm{var}(e)\cap(\bar v_{i+1}\cup\dots\cup\bar v_n)=\emptyset \},
  \end{aligned}
\]
where $R$ consists of the filters, the \genop{} expressions and each \code{OPTIONAL}, \code{MINUS} or \code{BIND} element~$e$ that shares a variable with $\bar v_1\cup\dots\cup\bar v_n$ or with an element~$e'$ preceding~$e$ such that $e'\in R$ or $\mathrm{var}(e')\cap(\bar v_1\cup\dots\cup\bar v_n)\neq\emptyset$. The translation of the group graph pattern is built from the algebra expressions $T_0,\dots,T_n$. Each $T_i$ joins~$Q_i$ with the solutions of~$T_{i-1}$ extended by~$G_i$:
\[
  T_0 = Q_0, \qquad T_i = Q_i \bowtie_{\theta_i} \genop_{\theta_i}(p_i,\bar v_i,m_i,T_{i-1}) \quad (1\le i\le n).
\]
As in SPARQL~1.1~\cite{sparql11query}, the elements of each~$Q_i$ translate in written order, and $\bowtie_\theta$ is the similarity join with the threshold~$\theta$. In Figure~\ref{fig:running}(b), $Q_0$ is empty and $Q_1$ consists of \code{?player a ex:Athlete}, so the group translates to $Q_1 \bowtie_{0.85} \genop_{0.85}(p_1,(\mathord{?}\mathit{player}),m_1,Q_0)$. The \code{OPTIONAL}, \code{MINUS} and \code{BIND} elements of~$R$ are then applied to~$T_n$ in written order. The filters are applied to the result as in SPARQL. The resulting expression yields the solutions of the group.

Following P\'erez et al.~\cite{sparql11}, we define the set $\sem{E}$ of solutions of an algebra expression~$E$ over the queried graph. For \genop{},
\[
  \begin{aligned}
  & \sem{\genop_\theta(p,\bar v,m,P)} = \{ \mu \cup [\bar v \mapsto \bar w]|_{\bar v\setminus\mathrm{dom}(\mu)} \,{:}\, \mu \in \sem{P},\\
  & \qquad \mathrm{ph}(p)\subseteq\mathrm{dom}(\mu), \bar w \in \mathrm{LLM}_m(p[\mu]), \mu\approx_\theta[\bar v\mapsto\bar w] \}.
  \end{aligned}
\]
Here $\mathrm{LLM}_m(q)$ is the set of tuples $\bar w=(w_1,\dots,w_k)$ of string literals that the LLM~$m$ returns for a prompt~$q$, with one literal for each output variable, and $[\bar v\mapsto\bar w]$ assigns~$w_j$ to each $\mathord{?}v_j$. If $\mu$ already binds an output variable $\mathord{?}v_j$, the condition $\mu\approx_\theta[\bar v\mapsto\bar w]$ of Section~\ref{sec:grounding} compares~$w_j$ with the labels of $\mu(\mathord{?}v_j)$, and the solution retains the term of~$\mu$. The set $\mathrm{ph}(p)$ contains the variables $\mathord{?}x$ with a placeholder \code{\{?x\}} in~$p$. For an RDF term~$t$, $\mathrm{str}(t)$ is the lexical form of~$t$ if $t$ is a literal and $t$ itself if $t$ is an IRI. The prompt $p[\mu]$ replaces the placeholder \code{\{?x\}} of each variable $\mathord{?}x\in\mathrm{ph}(p)$ by $\mathrm{str}(\mu(\mathord{?}x))$.

In the context-mode \genop{} of Figure~\ref{fig:running}(a), $\mathrm{ph}(p)=\{\text{\code{?name}}\}$, and the prompt \code{Position of \{?name\}?} becomes \code{Position of Lionel Messi?} for the solution of~$P$ that binds \code{?name} to \code{"Lionel Messi"}. The LLM answers \code{Forward}, so the extended solution binds \code{?pos} to \code{"Forward"}. In base mode, $\mathrm{ph}(p)=\emptyset$ and $p[\mu]=p$ for every~$\mu$, so every generated tuple extends every solution of~$P$. In Figure~\ref{fig:running}(b), the only solution of~$P$ is the empty solution, so every generated name forms one solution of \genop{}.

A context-mode \genop{} in the part~$B$ of a pattern $A$ \code{OPTIONAL} \{$B$\} can refer to variables of~$A$ in its placeholders. \gensparql{} therefore evaluates~$B$ separately for each solution~$\mu$ of~$A$, with $Q_0$ of the group of~$B$ joined with~$\{\mu\}$, and keeps~$\mu$ unextended if the evaluation yields no solution.

\subsection{Mapping Generated Names to Entities}
\label{sec:grounding}

The similarity join $\bowtie_\theta$ maps a generated name to an entity of the graph if the similarity of the name to a label of the entity is at least~$\theta$. For an IRI~$t$, $\mathrm{labels}(t)$ is the set of values of~$t$ for the first property among \code{rdfs:label}, \code{skos:prefLabel}, \code{schema:name} and \code{foaf:name} for which $t$ has a value in the queried graph. If $t$ has no value for these properties, $\mathrm{labels}(t)$ consists of the local name of~$t$. For a literal~$t$, $\mathrm{labels}(t)=\{\mathrm{str}(t)\}$. Each query selects the similarity $\mathrm{sim}$ of two strings from the cosine similarity of text embeddings~\cite{embsim} and the Jaccard similarity of lowercase word sets.

A solution~$\mu_1$ and a solution~$\mu_2$ that binds the output variables~$\bar v$ of a \genop{} are \emph{$\theta$-compatible}, written $\mu_1\approx_\theta\mu_2$, if every shared variable $\mathord{?}x\in\mathrm{dom}(\mu_1)\cap\mathrm{dom}(\mu_2)$ satisfies
\[
  \begin{cases}
  \max_{\ell\,\in\,\mathrm{labels}(\mu_1(\mathord{?}x))} \mathrm{sim}\bigl(\mu_2(\mathord{?}x),\ell\bigr) \ge \theta & \text{if } \mathord{?}x\in\mathrm{gen}(\mu_2),\\
  \mu_1(\mathord{?}x)=\mu_2(\mathord{?}x) & \text{otherwise,}
  \end{cases}
\]where $\mathrm{gen}(\mu_2)$ is the set of variables that $\mu_2$ binds to literals generated by \genop{}. For the solutions $\Omega_1$ of $Q_i$ and $\Omega_2$ of \genop{}, the similarity join is
\[
  \Omega_1 \bowtie_\theta \Omega_2 = \{ \mu_1 \cup \mu_2|_{\mathrm{dom}(\mu_2)\setminus\mathrm{dom}(\mu_1)} \,{:}\, \mu_1\in\Omega_1, \mu_2\in\Omega_2, \mu_1\approx_\theta\mu_2 \},
\]
where $\mu_2$ is restricted to the variables unbound in~$\mu_1$, so every shared variable retains the term of~$\mu_1$. For an output variable $\mathord{?}x\in\bar v$ of a \genop{} with the threshold~$\theta$ and a variable or RDF term~$t$, let $\mu$ be a solution with a generated literal $\mu(\mathord{?}x)$ and a defined $\mu(t)$ where $\mu(t)=t$ if $t$ is an RDF term. Under~$\mu$, the expression ${\mathord{?}x\mathrel{\texttt{=}}t}$ evaluates to true if the similarity of~$\mu(\mathord{?}x)$ to a label of~$\mu(t)$ is at least~$\theta$ and to false otherwise, and ${\mathord{?}x\mathrel{\texttt{!=}}t}$ evaluates to the opposite value. All other joins and expressions follow SPARQL~\cite{sparql11query,sparql11}.

In Figure~\ref{fig:running}(b), the similarity join with $\theta=0.85$ and the Jaccard similarity yields the mapping shown to the right of the query. Every IRI that the similarity join binds comes from a solution of~$Q_i$, so every mapped entity belongs to the graph. A name with similarity at least~$\theta$ to labels of several entities is mapped to each of them.

\subsection{Query Planning and Implementation}
\label{sec:implementation}

The planner orders the elements of a group as in the translation of Section~\ref{sec:genop}, so \gensparql{} sends one prompt for every solution of~$P$ in context mode and a single prompt in base mode. Filters without \code{EXISTS} over variables bound by the triple patterns of~$Q_0$ precede~$G_1$ and remove solutions before the prompts. In Figure~\ref{fig:running}(a), the LLM is invoked for the 26 Argentine players out of the 1,248 players in the graph, wherever \genop{} stands in the group.

\gensparql{} extends the query engine ARQ of Apache Jena~\cite{jena} with a parser for the syntax and compiles a query into the algebra of Section~\ref{sec:genop}. The engine evaluates \genop{}, the similarity join, \code{FILTER} and every \code{OPTIONAL}, \code{MINUS} or \code{UNION} that contains \genop{}, and ARQ evaluates the other operators. \gensparql{} sends prompts to the Anthropic API, to OpenRouter and to OpenAI-compatible APIs, including those of the inference servers vLLM and Ollama for self-hosted models. The prompts request the generated tuples as a JSON array that is empty if the LLM has no answer. \gensparql{} obtains embeddings from OpenAI-compatible APIs and falls back to the Jaccard similarity if the embedding model fails. The default threshold~$\theta_0$ is an engine parameter with the value 0.85. An HTTP service embeds the SPARQL server Fuseki and serves the workbench.

\section{Demonstration}
\label{sec:demo}

Attendees use \gensparql{} through a web workbench served by the HTTP service. The workbench offers the World Cup graph of Figure~\ref{fig:kg} (1,248 players, 48 teams) and a graph of 1,018 Nobel laureates. Attendees connect the workbench to an LLM and an embedding model through OpenAI-compatible endpoints and choose the embedding or Jaccard similarity and the threshold~$\theta$. The workbench shows the answers, the number of LLM calls and a mapping view that connects each generated name with its matched entities.
\begin{enumerate}
\item \emph{Generating attributes.} Attendees will use the chosen LLM to generate attributes absent from the graph, such as the positions in Figure~\ref{fig:running}(a) or the birth years of the Nobel laureates. The workbench compares the values generated by the three LLMs of Section~\ref{sec:eval} with Wikidata.
\item \emph{Mapping names to entities.} Attendees will list the national teams of the World Cup or the fields of the Nobel Prize with a base-mode \genop{} as in Figure~\ref{fig:running}(b) and vary the threshold~$\theta$. With both similarities at $\theta=0.85$, the similarity join maps neither United States nor Physics, because the graphs label the two entities USA and Nobel Prize in Physics. In the mapping view of the national teams, $\theta=0.8$ with the embedding similarity adds only incorrect matches such as Australia to USA.
\item \emph{Counting LLM calls.} Attendees will ask for the players of Group~A with the generated position defender or for the contributions of the Physics laureates since 2020. The workbench replays each query step by step in the order of the planner and with \genop{} evaluated first, and shows the LLM calls of both orders (Section~\ref{sec:eval}).
\item \emph{Writing new queries and uploading graphs.} Attendees will edit sample queries or write \genop{} queries with prompts and patterns of their choice. Attendees can also upload a graph of their own as a Turtle file.
\end{enumerate}

\paragraph{Live and recorded modes.}
In the live mode, the service evaluates each query with the chosen LLM. At the venue, we will run the service, the workbench and Ollama with Mistral-7B and bge-large-en-v1.5 on one laptop, so the live mode works without Internet access. With Internet access, attendees can also choose remote endpoints such as a vLLM server or a hosted LLM provider. The recorded mode on the project page replays the queries of (1)--(3) with the answers of Qwen3-30B, Qwen3-8B and Mistral-7B from Section~\ref{sec:eval} and replaces the live mode at the venue if the live system fails. The workbench also runs from the repository with a local LLM and without a provider account.

\section{Evaluation}
\label{sec:eval}

We built the World Cup graph and the Nobel graph from Wikidata without the positions of the players and the birth years of the laureates. We evaluated \gensparql{} on both graphs with the open-weight LLMs Qwen3-30B-A3B-Instruct-2507, Qwen3-8B in non-thinking mode and Mistral-7B-Instruct-v0.3 at temperature~0 in the inference server vLLM. The embedding similarity used the model bge-large-en-v1.5. The repository contains the queries, the gold standard from Wikidata and the results.

\paragraph{Generating attributes.}
Two queries with a context-mode \genop{} prompted each LLM for the position of each player and the birth year of each laureate. The position prompt asks for one of the groups goalkeeper, defender, midfielder and forward. The gold standard maps the positions in Wikidata to the same groups and holds a group for 1,119 players and a birth year for 990 laureates. An answer is correct if the answer names a single group or year that the gold standard lists for the entity. Table~\ref{tab:eval} reports the share of answered entities and the share of correct answers among the answered entities in the gold standard. Two baselines for the positions answer midfielder for every player or guess uniformly among the groups and are correct for 33\% and 26\% of the players. The LLMs are correct for 55--59\% of the positions and 69--90\% of the birth years. Empty arrays for 25\% of the players lower the answer rate of \mbox{Qwen3-8B}.

\begin{table}[htbp]
\centering
\caption{Results for the generated attributes and names.}
\label{tab:eval}
\small
\setlength{\tabcolsep}{3pt}
\begin{tabular}{@{}lcccccc@{}}
\toprule
& \multicolumn{2}{c}{Position} & \multicolumn{2}{c}{Birth year} & \multicolumn{2}{c}{Generated names} \\
\cmidrule(lr){2-3}\cmidrule(lr){4-5}\cmidrule(l){6-7}
LLM & answered & correct & answered & correct & total & not in graph \\
\midrule
Qwen3-30B & 99.9\% & 58.4\% & 97.4\% & 90.2\% & 71 & 20\% \\
Qwen3-8B & 74.7\% & 54.7\% & 95.7\% & 68.8\% & 44 & 61\% \\
Mistral-7B & 99.9\% & 58.7\% & 99.9\% & 89.8\% & 113 & 35\% \\
\bottomrule
\end{tabular}
\end{table}

\paragraph{Mapping names to entities.}
Six base-mode queries over the two graphs request lists of entities such as the teams of the World Cup and the fields of the Nobel Prize. As in Figure~\ref{fig:running}(b), the similarity join of each query maps the generated names to the solutions of one triple pattern. Between 20\% and 61\% of the generated names of each LLM denote no entity of the graph (Table~\ref{tab:eval}). The names of the three LLMs denote 145 entities of the graph, counted once per LLM. Precision and recall compare the mapped entities with the 145 entities. With the Jaccard similarity, the precision is 1 for $\theta\ge0.6$ and the recall at $\theta=0.85$ is 67\%, because name variants such as Emiliano Martinez remain unmapped. The embedding similarity maps such variants and has a recall of 83\% with one incorrect match at $\theta=0.85$ and of 79\% without an incorrect match at $\theta=0.9$. At lower thresholds, the precision of the embedding similarity decreases to 0.91 at $\theta=0.8$, 0.64 at $\theta=0.7$ and 0.07 at $\theta=0.5$.

\paragraph{Counting LLM calls.}
For the query for the defenders of Group~A and the query for the contributions of the Physics laureates since 2020 (Section~\ref{sec:demo}), we counted the LLM calls in the order of the planner and with \genop{} evaluated first. The planner evaluates the patterns for Group~A and the \code{FILTER} on the award year before \genop{}. The LLM is therefore called once for each of the 104 players of Group~A instead of the 1,248 players in the graph and once for each of the 17 Physics awards since 2020 instead of the 230 Physics awards. The numbers of calls do not depend on the LLM.

\section{Conclusion}
\label{sec:conclusion}

We presented \gensparql{}, an extension of SPARQL with the operator \genop{}. \gensparql{} answers queries with values generated by an LLM, maps generated names to entities of the graph and restricts LLM invocations to the bindings selected by the other graph patterns and filters. Every mapped entity belongs to the graph, whereas the correctness of a generated value depends on the LLM. Quantifying the uncertainty of generated values and of their mapping to entities and propagating the uncertainty through the evaluation of queries with \genop{} remain open questions.

\section*{Ethical Use of Data}

The graphs contain public facts from Wikidata (CC0) and official records, including the citizenship and gender of the Nobel laureates. The evaluation involves no human participants, and the workbench collects no user data. Generated values about people can be incorrect or biased and are not stored in the graphs. Names of people in prompts and embedding requests reach external servers only when users select remote endpoints.

\begin{acks}
The project was funded by the Deutsche Forschungsgemeinschaft (DFG, German Research Foundation) - SFB~1574 - Project number 471687386 (``Circular Factory''). The authors thank the International Max Planck Research School for Intelligent Systems (IMPRS-IS) for supporting Jingcheng Wu, Yi Wang and Hongkuan Zhou.
\end{acks}

\bibliographystyle{ACM-Reference-Format}
\bibliography{references}

\end{document}